\documentstyle[twocolumn,psfig]{mn}

\newif\ifAMStwofonts
\AMStwofontstrue

\def\gsim{~\rlap{$>$}{\lower 1.0ex\hbox{$\sim$}}}

\def\ltsim{\lower.5ex\hbox{$\; \buildrel < \over \sim \;$}}
\def\gtsim{\lower.5ex\hbox{$\; \buildrel > \over \sim \;$}}
\def\ltsim{\lower.5ex\hbox{$\; \buildrel < \over \sim \;$}}
\def\gtsim{\lower.5ex\hbox{$\; \buildrel > \over \sim \;$}}

\def\rhoi{{\rho_{_{\rm I}}}}
\def\rhob{{\rho_{_{\rm b}}}}
\def\rhoa{{\rho_{_{\rm a}}}}
\def\gammai{{\gamma_{_{\rm I}}}}
\def\gammaa{{\gamma_{_{\rm a}}}}
\def\gammab{{{\gamma_{_{\rm b}}}}}
\def\ui{{u_{_{\rm I}}}}
\def\ub{{u_{_{\rm b}}}}
\def\ua{{u_{_{\rm a}}}}

\def\dd{\,{\rm d}}

\def\gammab{\gamma_{_{\rm b}}}

\def\rhob{{\rho_{_{\rm b}}}}

\newcommand{\etal}{{\it et. al.}\ }

\begin{document}
\textheight 9in
\title[A porous intracluster medium]{
Hydrostatic equilibrium of a porous intracluster medium: implications for
mass fraction and X-ray luminosity}

\author[Nusser \&  Silk ] { Adi Nusser$^{1}$ and  Joseph Silk$^{2}$ 
 \\\\
  $^{1}$Physics Department- Technion, Haifa 32000, Israel\\
  $^{2}$Astrophysics, Oxford University, Keble Road, Oxford OX1 3HR, UK}

\maketitle

\begin{abstract}
The presence of dilute hot cavities in the intracluster medium (ICM) 
at the cores of clusters of galaxies changes the 
relation between gas temperature and its X-ray emission properties. 
Using the hydrostatic equations of a porous medium 
we solve for the ICM density for a given temperature as a function 
of the filling factor of dilute bubbles. 
We find that  at a given temperature, 
the  core X-ray luminosity increases with the filling factor. 
If the frequency of AGNs in clusters were higher in the past, then
the filling factor could  correspondingly be significant, with implications 
for the cluster scaling relations at high redshifts. This is especially important for the core properties,
including the temperature-luminosity ($L_X-T$) relation and estimates of the core gas mass.   
The results imply an epoch-dependent sensitivity of the $L_X-T$ relation in the core to the porosity of the ICM. Detection of such an effect would give new insights into AGN feedback.

\end{abstract}
 
                           
\section{introduction}
X-ray observations of the intracluster medium (ICM) of massive clusters 
indicate the presence of cavities seen as depressions in the X-ray surface 
brightness (e.g. Boehringer \etal 1993; Fabian \etal 2000; McNamara \etal 2000; B\^irzan \etal 2004).  These cavities or bubbles 
 are believed to have been generated by 
AGN activity of central galaxies and possibly other galaxies in the cluster
(Nusser, Silk \& Babul 2006; Best et al. 2007; Martini, Mulchaey, \& 
Kelson 2007).

The observed properties of the x-ray emitting gas in clusters of galaxies
provide strong constraints on the thermodynamic properties of the ICM. The most  satisfactory model which accounts for the observed self-similar scaling properties of outer regions  invokes  galaxy formation by cooling gas (Suginohara \& Ostriker 1998; Lewis et al. 2000; Pearce et al. 2000; Dav\'e, Katz \& Weinberg 2002; Ettori et al. 2004; Nagai, Kravtsov \& Vikhlinin 2007), but this in turn leads to a major puzzle.
Approximately twice as much cold gas is predicted as compared with the mass in stars.
AGN feedback in the core can  resolve this problem (e.g. Quilis \etal 2001, Babul \etal 2002, Dalla Vecchia \etal 2004, Ruszkowski \etal 2004,  Voit \& 
Donahue 2005, Br\"uggen \etal 2005, Roychowdhury \etal 2005, 
Sijacki \& Springel 2006, Nusser, Silk \& Babul 2006, Peterson \& Fabian 2006, McNamara \& 
Nulsen 2007, Cattaneo \& 
Teyssier 2007). Related discussions of the role of AGN heating of the ICM are given in semi-analytic 
modelling of galaxy formation (Croton et al. 2006; Bower et al. 2006). We point out here that AGN feedback drives porosity in the ICM and modifies the hydrostatic properties of cluster cores. A higher AGN activity in the past 
leads to a   redshift evolution of the scaling relations.

\section{The hydrostatic equation for the ambient medium}
\label{sec:eom}

A  formalism for following the evolution of bubbles statistically
has  been developed in Nusser, Silk \& Babul (2006). 
In this formalism,
the hot dilute bubbles and the rest of the  ICM are a two-fluid system that is best described
as a single fluid which we term the {\it ambient medium}.
The   representation in terms of  a single medium 
is applicable only  if local pressure equilibrium
between the bubbles and the residual ICM is established. Hereafter, quantities related to the ambient medium, the bubbles, and the residual ICM (gas 
outside bubbles) are denoted by the subscripts, $a$, $b$, and $I$, respectively. 
Suppose that the volume filling factor of bubbles at $r$ is $F(r)$.
The  ambient density, $\rhoa$, at position $r$,  is defined as
\begin{equation}
\rhoa=(1-F)\rhoi+F\rhob \; ,
\label{eq:rhoadef}
\end{equation}
in terms of the density inside bubbles, $\rhob$, and the density of the ICM, $\rhoi$, also at $r$.
The local ambient energy, $\ua$, per unit mass as
\begin{equation}
\ua=\frac{(1-F)\rhoi\ui+F\rhob\ub}{\rhoa} \; ,
\label{eq:uadef}
\end{equation}
where $\ub$ and $\ui$ are the energy per unit mass of the material inside the bubbles 
and of residual the ICM.
If the pressure, $p$, is 
$p=(\gamma-1)u\rho$ for both the ICM and the material inside the bubbles,
 The equation of state of the ambient 
medium is 
\begin{equation}
p=(\gammaa-1)\rhoa\ua\; ,
\end{equation}
where 
\begin{equation}
\gammaa-1=\frac{(\gammai-1)(\gammab-1)}{(1-F)(\gammab-1)+F(\gammai-1)}\; ,
\end{equation} 
and we have assumed local pressure equilibrium  between the bubbles and the ICM.
Hydrostatic equilibrium of  the ambient medium in a gravitational field $g(r)$ is described by, 
\begin{equation}
0= g -\frac{1}{\rhoa}\frac{\dd  p}{\dd r}\; .
\label{eq:euler}
\end{equation}

We numerically integrate these equations assuming NFW profile
with a virial radius of $4\; \rm Mpc$ and a concentration parameter of $7$ (Neto et al. 2007).
We assume that the gas outside the bubbles has a temperature of $6\times 10^7\rm K$ and that 
the core baryonic to dark mass ratio equals 0.14. We take a core radius, $r_{\rm c} =100 \; \rm kpc$. 
This choice of the parameters yield reasonable 
temperature and density in the core. Of course, 
the choice is not unique. Once can obtain similar characteristic of the central region for 
smaller virial radii at the expense of changing the concentration and the core radius. For simplicity we assume that the mass in the bubbles is negligible.
\section{results}
\begin{figure*}
\centering
%\begin{sideways}
\mbox{\psfig{figure=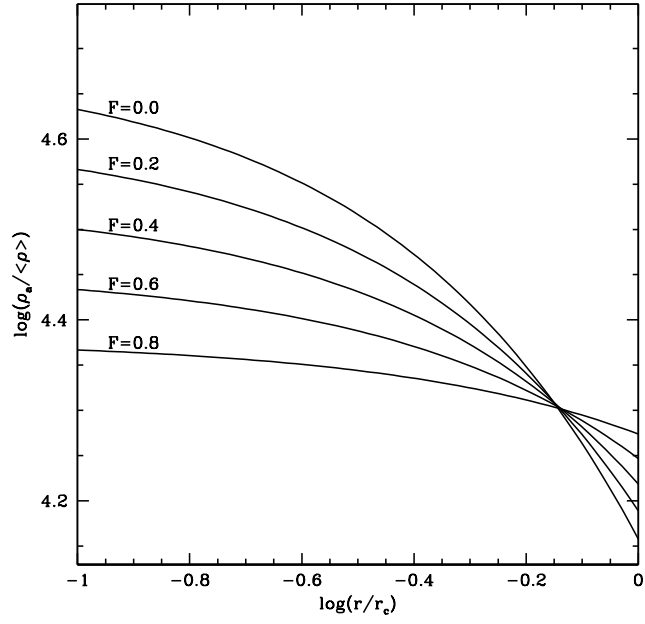,height=3.5in}}
%\end{sideways}
\vspace{0.25cm}
\caption{The ambient density, $\rhoa$, profiles (in units of the background value) 
as a function of radius (in units of $r_{\rm c}$, for several values of the bubble filling factor. }
\label{fig:fig1}
\end{figure*}

\begin{figure*}
\centering
%\begin{sideways}
\mbox{\psfig{figure=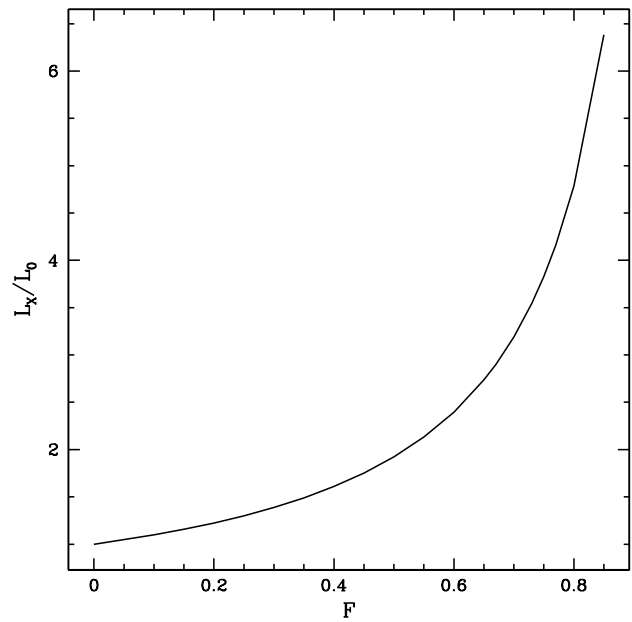,height=3.5in}}
%\end{sideways}
\vspace{0.25cm}
\caption{The normalized X-ray luminosity versus the filling factor.}
\label{fig:fig2}
\end{figure*}

The density averaged over the bubbles and residual core gas
in the ambient medium, $\rhoa$,  as a function of radius is 
plotted in figure (\ref{fig:fig1}) for various values of the filling factor. In each case, a constant filling factor is assumed. 
The profiles become flatter as  the filling factor is increased. 
This can be seen analytically from the hydrostatic equation
(\ref{eq:euler}) as follows. Substituting  $\rhoa=(1-F) \rhoi$ and 
 $P=(\gamma -1) \rhoi \ui$ in this equation we obtain, for $\ui=const$ and $F=const$,  
 \begin{equation}
 g=(\gamma -1) \ui \frac{\dd \ln \rhoi^{1/(1-F)}}{\dd r} \; . 
 \end{equation}
This implies  that the density $\rhoi(F)$ as a function of the filling factor 
scales like $\rhoi(F)\sim\rhoi(F=0)^{1-F}$ which in turn leads to a similar scaling for 
the ambient density plotted in the figure.
Since $0\leq F<1$, this scaling yields flatter profiles for 
larger $F$.   

The modification to   the density profile causes a change in the corresponding 
core X-ray luminosity, $L_X\propto \int\dd r r^2 (1-F) \rhoi^2(r)=
\int\dd r r^2 (1-F)^{-1} \rhoa^2(r) $.
The X-ray luminosity (normalized to the value for $F=0$)  is plotted
in figure  (\ref{fig:fig2}). 
The luminosity is an increasing function of the filling factor.  
This is due  essentially to
 the larger ambient  density in the outer regions 
for a porous medium, amplified by the $(1-F)^{-1}$ factor in the expression
for $L_X$. 
The net result is a boost in the core X-ray luminosity.

\section{Discussion}
\label{sec:sum}

\begin{table*}
 \centering
 \begin{minipage}{140mm}
  \caption{Filling factors of radio lobes in a few 
  rich clusters (data taken from B\^irzan et al. 2004).}
  \begin{tabular}{@{}lllllll@{}}
                   & Perseus     & Hydra A   & Cygnus A  & MKW 3S    &  RBS 797     & A2199 \\
 F($R<30\rm kpc$)  & 0.006 -0.036&  0.07-0.14& 0.11-0.25 & 0.12-0.23 &  0.023 -0.14& 0.006-0.03\\
 F($R<40\rm kpc$)  & 0.002-0.015 &  0.06-0.11& 0.12-0.26 & 0.09-0.19 &  0.01 -0.066&  0.002-0.01 
\end{tabular}
\end{minipage}
\end{table*}

Our results may be interpreted as follows: 
\begin{enumerate}
\item For a given observed X-ray emission and spectroscopic temperature of 
the residual ICM gas (outside the bubbles), hydrostatic 
equilibrium can yield erroneous mass estimates if the porosity of the medium is 
not properly included. 
\item  The $L_X-T$ relation  will depend on the porosity.   The effect is enhanced if 
analysis is restricted to the core where the porosity is highest. 
This is in addition to  changes resulting  from gas  cooling in the core (Allen \& Fabian 1998,  Markevitch
   1998). 
\item The ICM porosity is expected to be larger in the past  due to an enhanced frequency of AGN activity. This leads to redshift dependence of a core  $L_X-T$ relation
which is different from that seen for the outer parts of the cluster (Kotov \& Vikhlinin 2005).  
 \item Porosity enhances the X-ray emission and hence shortens the 
cooling time. This leads to more AGN activity and the effect is only quenched 
when the porosity becomes sufficiently large, when 
only little amount of gas remains available for 
feeding the AGN. At this point the cold gas 
is all driven out of the core until there is time for a new   
episode of cooling and associated AGN activity.
\end{enumerate}

Predictions of the modified hydrostatic equation
should be compared  against observations 
by the analysis of hydrostatic equilibrium of clusters. Porosity of individual clusters is needed 
for such a study. But reliable estimates of the 
porosity is not always possible using current data. 
One could alternatively proceed by 
solving the modified equation for assumed 
porosity as a function of cluster temperature and mass. The scatter in the relevant relations could then be 
minimised with respect to the assumed porosity values. 

Rough estimates of porosity 
could be derived from  the results of B\^irzan et al. (2004) (see also Dunn \& Fabian 2004 and Dunn, Fabian \& Taylor 2005, Dunn \& Fabian 2006).
Using their table 2 for the sizes of prominent active bubbles (radio lobes),
we compute the filling factors corresponding to some of their clusters.
The filling factors  
within a distance  of $30$ and $40\rm kpc$ from the center are listed in table 1 here. 
The range of the filling factors corresponds to the uncertainty in projection effects (cf. B\^rzan et al.). Each cluster in  table 1 contains 2 radio 
lobes and the upper (lower) limiting value of the  filling factor is obtained by summing the maximum (minimum)
values of the sizes of both lobes.  
Table 1 demonstrates that the porosity resulting from active prominent bubbles 
could be significant.   
Note that, low contrast bubbles remain undetected (Dunn \& Fabian 2006) so that the porosity could be much higher than the 
estimate based on observed bubbles. 
Therefore, more rigorous estimates of the porosity, especially at larger radii and at 
different redshifts,  by detailed 
observational studies of higher S/N are needed in order to test the evolutionary implications of the AGN feedback hypothesis. 

 The modified hydrostatic equation is valid for 
any two-component fluid   where the density of one of the
components  is negligible. The generalization to multi-component fluids of any densities could easily 
be done using the derivation  described in
Nusser, Silk \& Babul (2006).

\section{Acknowledgments}
 A.N.  would like
to express his  gratitude  for the  hospitality of the  Astrophysics Department at the University of Oxford.
This research is supported by the German-Israeli Foundation 
for Development and Research and a grant from the 
Technion Asher Space Research Institute.  

\protect
\end{document}